\documentclass{ws-rv975x65}
\usepackage{subfigure}     
\usepackage{ws-rv-thm}     
\usepackage{ws-rv-van}     

\RequirePackage{lineno}
\usepackage{graphicx}
\usepackage{mathrsfs}
\usepackage{amsmath}
\usepackage{amssymb,amsfonts}
\usepackage[x11names]{xcolor}

\makeindex

\begin{document}

\chapter[AGN as High-Energy Neutrino Sources]{Active Galactic Nuclei as High-Energy Neutrino Sources}\label{ra_ch2}

\author[Kohta Murase]{Kohta Murase\footnote{murase@psu.edu}}
\address{Center for Particle and Gravitational Astrophysics; Department of Physics; Department of Astronomy \& Astrophysics, 
The Pennsylvania State University, University Park, Pennsylvania 16802, USA}
\address{Institute for Advanced Study, Princeton, New Jersey 08540, USA}

\begin{abstract}
Active galactic nuclei (AGN) are believed to be promising candidates of extragalactic cosmic-ray accelerators and sources, and associated high-energy neutrino and hadronic gamma-ray emission has been studied for many years.  
We review models of high-energy neutrino production in AGN and discuss their implications for the latest IceCube observation of the diffuse neutrino intensity.  
\end{abstract}
\body

\section{Introduction}\label{ra_sec1}
Active galactic nuclei (AGN), which are powered by accretion of mass onto supermassive black holes (SMBHs) at the center of their host galaxies and/or the rotational energy of SMBHs, are the most luminous persistent sources of electromagnetic radiation in the Universe.  They have also been of interest as powerful high-energy cosmic-ray accelerators (CRs), including ultrahigh-energy cosmic rays (UHECRs).  High-energy neutrinos from AGN have been discussed since the late 70s at least~\cite{ber77,euc77,brs90}. If protons are accelerated by e.g., the diffusive shock acceleration mechanism, because the optical and X-ray radiation density is rather high in the vicinity of a SMBH, the CRs may efficiently interact with the ambient photons. Early models that attempted to interpret X-ray emission with pair cascades~\cite{ke86} led to very large diffuse neutrino intensities~\cite{Stecker:1991vm,Szabo:1994qx,AlvarezMuniz:2004uz}.  The X-ray origin is now thought to be Comptonized disk emission~\cite{hm91} and the originally predicted fluxes have been largely constrained by neutrino observations themselves~\cite{Achterberg:2005fs}.  However, these early models motivated the searches for cosmic high-energy neutrinos with water- or ice-Cherenkov detectors.   

Both observational and theoretical multi-wavelength efforts have helped us understand the physics of AGN, which also lead to different proposals for CR acceleration and associated neutrino production in AGN.  In particular, radio-loud AGN and their on-axis objects ``blazars'' are most widely discussed as powerful non-thermal sources~\cite{Dermer:2012ry}. Our gamma-ray view of blazars has been drastically enriched by EGRET on the {\it \it Compton Gamma ray Observatory}, {\it Fermi}, and various ground-based Cherenkov telescopes. They are the dominant sources in the extragalactic gamma-ray sky~\cite{Inoue:2014ona}, which has tempted us to speculate that the radio-loud AGN including blazars are powerful accelerators of protons as well as electrons~\cite{bs87,mkb91,m93,Aharonian:2000pv,Mucke:2000rn,Muecke:2002bi}.  The origin of gamma-ray emission is still under debate even in the {\it Fermi} era. The standard explanation is inverse-Compton emission by non-thermal electrons (leptonic scenario)~\cite{Ghisellini:2009fj,Boettcher:2013wxa,Cerruti:2013dga,Dermer:2013cfa}, but lepto-hadronic scenarios have also been exploited to explain BL Lacertae objects (BL Lacs)~\cite{Boettcher:2013wxa,Cerruti:2014iwa,Yan:2014gma}, quasar-hosted galaxies (QHBs) including flat-spectrum radio quasars (FSRQs)~\cite{Atoyan:2002gu,Dermer:2012rg,Diltz:2015kha,Petropoulou:2015swa}, Fanaroff-Riley I (FR I) and Fanaroff-Riley II (FR II) radio galaxies~\cite{Atoyan:2008uy,Kachelriess:2008qx,Dermer:2008cy,Sahu:2012wv,Fujita:2015xva}.  A fraction of BL Lacs, so-called extreme BL Lacs, show very hard gamma-ray spectra, which could be explained by hadronic cascade emission induced by CRs propagating in intergalactic space~\cite{Essey:2009zg,Essey:2010er,Murase:2011cy,Takami:2013gfa,Yan:2014vsd}. 

The IceCube's discovery of cosmic high-energy neutrinos~\cite{Aartsen:2013bka,Aartsen:2013jdh,Aartsen:2014gkd,Aartsen:2015knd,Aartsen:2015rwa} raises new questions about the non-thermal properties of AGN.  Is observed gamma-ray emission produced by high-energy CRs accelerated in blazars and radio galaxies? Do AGN make a dominant contribution to the observed diffuse neutrino intensity? Are they the sources of UHECRs?  In this article, we discuss possibilities of neutrino production in AGN, with a focus on recent studies in light of the IceCube data.  AGN, especially radio-loud AGN, have been excellent targets of multi-wavelength observations. We will see that they are also promising targets of the multi-messenger astronomy.

\section{Models of AGN Neutrino Emission}
It is convenient to divide AGN into radio-quiet and radio-loud AGN.  In the radio-loud objects, the emission contribution from jets and bubbles or lobes is prominent especially at radio wavelengths. In the radio-quiet objects, the continuum emission comes from core regions within $\sim1-100~(GM_{\rm BH}/c^2)$ since jet and jet-related emission are weak. 
We first consider neutrino production in CR accelerators.  The accelerators can be AGN jets of radio-loud AGN, or AGN cores of both radio-loud and radio-quiet AGN.  Next, we consider the fate of CRs escaping from accelerators, and discuss neutrino production in CR reservoirs or during CR propagation. 

The diffuse neutrino intensity from extragalactic AGN is given by~\cite{Murase:2014foa} 
\begin{eqnarray}
\Phi_\nu&=&\frac{c}{4\pi H_0}\int^{\rm z_{\rm max}} \!\!\! dz \, \frac{1}{\sqrt{{(1+z)}^3\Omega_m+\Omega_\Lambda}} \int \!\!\! dL_\nu \, \frac{d n_s}{dL_\nu}(z) \frac{L_{E'_\nu}}{E'_\nu},
\end{eqnarray}
where $d n_s/dL_\nu$ is the neutrino luminosity function of the sources (per comoving volume per luminosity) and $z_{\rm max}$ is the maximum value of the redshift $z$ for a given source class. To make model predictions for the diffuse neutrino intensity, it is necessary to relate the calculated neutrino luminosity to the observed photon luminosity at some energy band.  In addition, one needs to normalize the CR spectrum. It is ideal to calculate CR acceleration from first principles, but our present knowledge on particle acceleration is not sufficient.  Phenomenologically, the neutrino flux can be normalized by the observed CR data or by the existing gamma-ray data. Or, one can introduce a phenomenological parameter such as the CR loading factor ($\xi_{\rm cr}$) to represent the efficiency of CR proton acceleration. 

\begin{figure}[t]
\begin{center}
\includegraphics[width=3.00in]{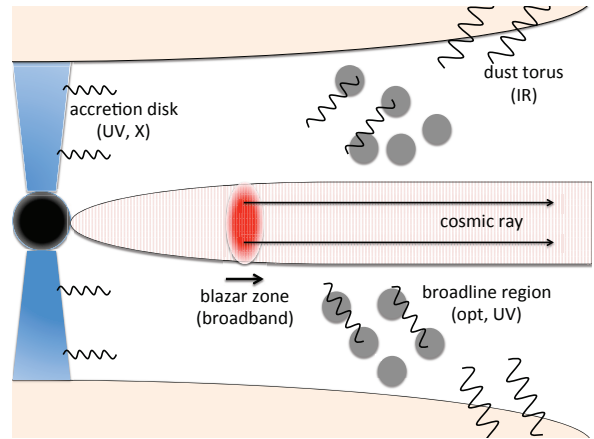}
\caption{Schematic picture of photohadronic interactions by CRs in inner jets of radio-loud AGN~\cite{Murase:2014foa}. 
\label{fig1}
}
\vspace{-1.\baselineskip}
\end{center}
\end{figure}

\subsection{AGN Jets}
The most promising site of non-thermal AGN emission is the jet of radio-loud AGN (see Fig.~1). Their broadband emission has been studied at multi-wavelengths from radio to gamma rays.  In particular, spectral energy distributions of blazars have been measured and modeled for many years.  There are two main blazar sub-classes, namely BL Lacs and QHBs (mostly FSRQs).  They differ mostly in their optical spectra, and FSRQs display strong broad emission lines, whereas BL Lacs are characterized by optical spectra showing at most weak emission lines or absorption features.  Continuum radiation of both blazar classes (FSRQs and BL Lacs) typically consists of two humps (see the left panel of Fig.~2).  The low-energy hump (peaking in the infrared to soft X-ray band) is well explained by synchrotron radiation from non-thermal electrons. The high-energy hump is conventionally attributed to inverse-Compton emission.  The spectral energy distributions of high-energy-peaked BL Lac objects (HBLs) are interpreted as synchrotron and synchrotron self-Compton components.  In contrast, those of low-energy-peaked BL Lac objects (LBLs) and FSRQs are generally well fit with synchrotron and external inverse-Compton components.  External radiation fields are naturally provided by the accretion-disk radiation, its scattered radiation from the broadline region (BLR), and infrared (IR) radiation from the dust torus surrounding a SMBH and the BLR (see Fig.~1).  Typical quasars show such broad optical and UV emission lines from the BLR, and the dust torus plays a key role in the AGN unification scheme~\cite{Antonucci:1993sg}. 

\begin{figure}[tb]
\begin{minipage}{.49\linewidth}
\includegraphics[width=\textwidth]{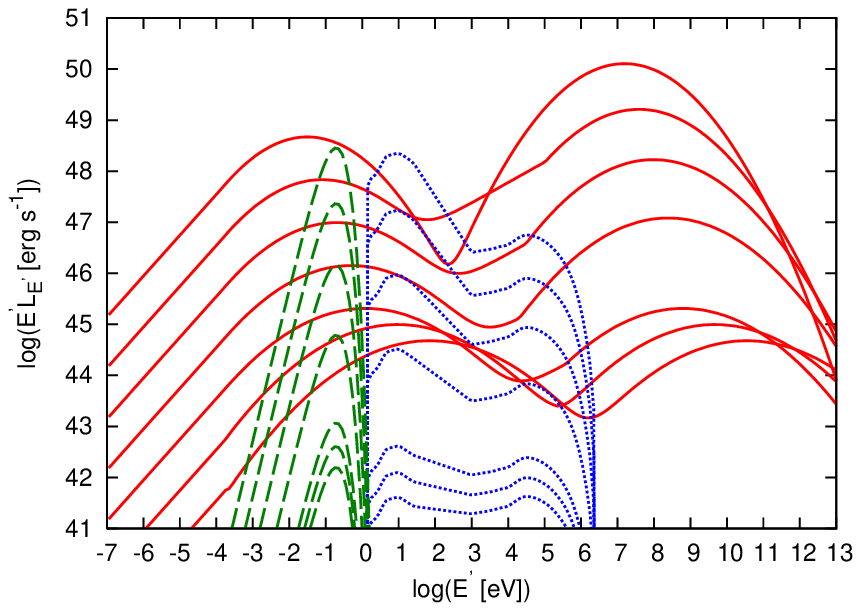}
\end{minipage}
\hfill
\begin{minipage}{.49\linewidth}
\includegraphics[width=\textwidth]{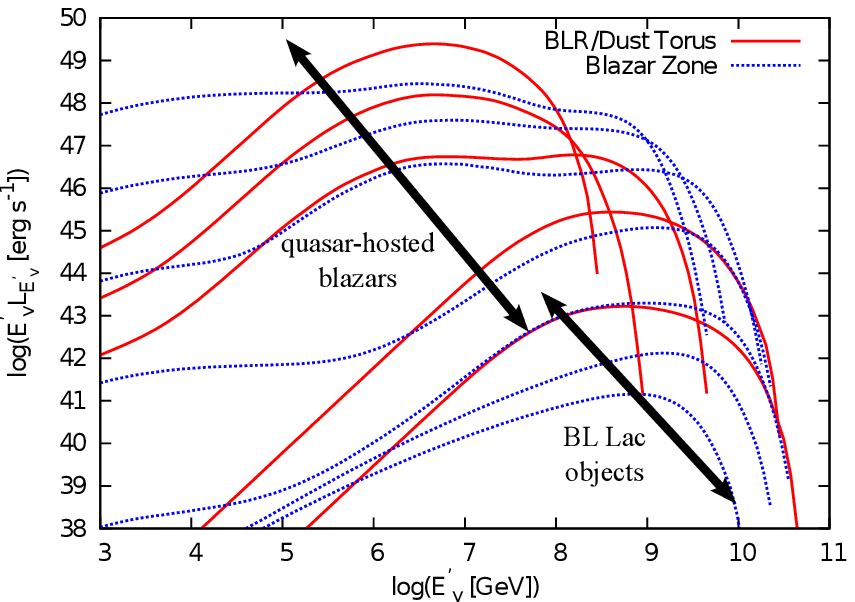}
\end{minipage} 
\caption{{\it Left panel}: Differential continuum luminosity spectra of observed photons from blazars~\cite{Murase:2014foa}. The sold, dotted, and dashed curves represent the non-thermal jet component, the accretion disk component, and the torus infrared component, respectively.  The radio luminosity at 5~GHz varies as $\log(L_{5{\rm GHz}})=$47, 46, 45, 44, 43, 42, and 41, in units of ${\rm erg}~{\rm s}^{-1}$ from top to bottom.
{\it Right panel}: Differential luminosity spectra of photohadronic neutrinos from blazars~\cite{Murase:2014foa}.  The muon neutrino spectrum is calculated for $s=2.0$ and $\xi_{\rm cr}=10$, with neutrino mixing.  From top to bottom, the radio luminosity varies corresponding to the left panel.
}
\end{figure}

High-energy protons may be accelerated by diffusive shock acceleration or stochastic acceleration in a jet.  
They interact with synchrotron photons provided by non-thermal electrons that are co-accelerated in jets~\cite{msb92,m95,Halzen:1997hw,Atoyan:2001ey,Muecke:2002bi,Dimitrakoudis:2013tpa}.  The effective optical depth to photomeson production is estimated to be
\begin{equation}
f_{p \gamma}(E'_p)\approx\frac{2 \kappa_\Delta \sigma_\Delta}{1+\beta} \frac{\Delta \bar{\varepsilon}_{\Delta}}{\bar{\varepsilon}_{\Delta}} 
\frac{L_{\rm rad}^s}{4 \pi r \Gamma^2 c E'_s} {\left(\frac{E'_p}{{E'}_p^s} \right)}^{\beta-1}, 
\end{equation}
where $\sigma_\Delta \sim 5 \times {10}^{-28}~{\rm cm}^2$, $\kappa_\Delta \sim 0.2$, $\bar{\varepsilon}_{\Delta} \sim 0.3$~GeV, $\Delta \bar{\varepsilon}_{\Delta} \sim 0.2$~GeV, $L_{\rm rad}^s$ is the jet synchrotron luminosity at the synchrotron peak $E'_s$, $r$ is the emission radius, $\Gamma$ is the bulk Lorentz factor of jets, and $\beta$ is the photon index of target photons.  Note that primes refer to quantities in the black-hole rest frame.
The characteristic energy of protons that interact with target photons with $E'_s$ is given by
\begin{equation}
{E'}_p^b\approx0.5\Gamma^2 m_p c^2 \bar{\varepsilon}_\Delta{(E'_s)}^{-1}.
\end{equation}
For BL Lac objects with $L_{\rm rad}^s\sim{10}^{45}~{\rm erg/s}~$ and $E'_s\sim10$~eV, we have
\begin{equation}
f_{p\gamma}(E'_p)\simeq7.8\times{10}^{-4}L_{\rm rad,45}^s \Gamma_1^{-4}{\delta t'}_5^{-1}{(E'_s/10~{\rm eV})}^{-1}
\left\{\begin{array}{ll}
{(E'_{\nu}/{E'}_{\nu}^{b})}^{\beta_h-1} 
& \mbox{($E'_p \leqq {E'}_{p}^{b}$)}
\\
{(E'_{\nu}/{E'}_{\nu}^{b})}^{\beta_l-1} 
& \mbox{(${E'}_{p}^{b} < E'_p$)}
\end{array} \right.
\end{equation}
where $\delta t'$ is the variability time in the black hole rest frame, and $\beta_l\sim1.5$ and $\beta_h\sim2.5$ are the low-energy and high-energy photon indices, respectively.  When cooling of mesons and muons is negligible, the characteristic neutrino energy corresponding to ${E'}_p^b$ is
\begin{equation}
{E'}_\nu^b\approx0.05{E'}_p^b\simeq80~{\rm PeV}~\Gamma_1^2{(E'_s/10~{\rm eV})}^{-1}.
\end{equation}
We immediately see the following features.  For a power-law CR spectrum such as $E_p^{-2}$, the resulting neutrino spectra should be hard since $f_{p\gamma}$ increases with energy.  As an example, let us consider BL Lacs, where external radiation fields are not relevant.  As shown in the right panel of Fig.~2, neutrino spectra of BL Lacs rise up to EeV energies, and the peak energy is much higher than $\sim1$~PeV and the Glashow resonance energy at 6.3~PeV.  Second, $f_{p\gamma}$ is quite sensitive to $\Gamma$.  This is one of the reasons why blazar neutrino models have large uncertainties in their predictions for the normalization of the neutrino flux.  

Next, we consider interactions with external photons provided by the BLR clouds and the IR dust torus.  The importance of BLR photons and IR photons for the neutrino production has been studied by several authors~\cite{Atoyan:2001ey,Atoyan:2002gu,Dermer:2012rg,Murase:2014foa}.  For the calculation, one can use empirical relations between the BLR/torus size and accretion-disk luminosity $L_{\rm AD}$~\cite{Ghisellini:2008zp,Kishimoto:2011hz}.  Then, assuming an isotropic distribution in the black hole rest frame, the photomeson production efficiency in the BLR is estimated to be~\cite{Murase:2014foa} 
\begin{equation}
f_{p\gamma}\approx \hat{n}_{\rm BL}\sigma_{p\gamma}^{\rm eff}r_{\rm BLR}\simeq5.4\times{10}^{-2}~f_{\rm cov,-1}L_{\rm AD,46.5}^{1/2},
\end{equation}
above the pion production threshold energy, where $f_{\rm cov}$ is the covering factor and $\sigma_{p\gamma}^{\rm eff}$ is the attenuation cross section of the photomeson production.  Similarly, the photomeson production efficiency for CR protons propagating in IR radiation fields supplied by the dust torus is estimated to be~\cite{Murase:2014foa} 
\begin{equation}
f_{p\gamma}\simeq0.89~L_{\rm AD,46.5}^{1/2}{(T_{\rm IR}/500~{\rm K})}^{-1},
\end{equation}
above the pion production threshold energy.  Importantly, $f_{p\gamma}$ does not depend on $\Gamma$ and $\delta t'$, which implies that the results on neutrino fluxes are much more insensitive to model parameters compared to the case of internal synchrotron target photon fields.  The photomeson production with external radiation fields is important and should not be neglected for luminous blazars such as LBLs and QHBs, leading to spectral bumps in the PeV and EeV range (see the right panel of Fig.~2).  Note that the accretion-disk emission is also important if $\tau_{\rm sc}\gtrsim f_{\rm cov}$.  

Final results of the diffuse neutrino intensity depend on the neutrino luminosity function.  It has been suggested that the spectral energy distributions of blazars evolve with luminosity, which is the so-called blazar sequence~\cite{Ghisellini:2008zp,Fossati:1998zn} (see the left panel of Fig.~2).  In the simple one-zone leptonic model, since $f_{p\gamma}$ increases with the observed photon luminosity, photohadronic interactions with broadline and IR emission in LBLs and QHBs play an important role~\cite{Murase:2014foa}.  As a result, the neutrino spectrum is roughly expressed by
\begin{equation}
E'_{\nu}L_{E'_\nu}\approx\frac{3}{8}{\rm min}[1,f_{p \gamma}](E'_{p}L_{E'_p})
\left\{\begin{array}{ll}
{(E'_{\nu}/{E'}_{\nu}^{b})}^{2} 
& \mbox{(for $E'_{\nu} \leqq {E'}_{\nu}^{b}$)}
\\
{(E'_{\nu}/{E'}_{\nu}^{b})}^{2-s} 
& \mbox{(for ${E'}_{\nu}^{b} < E'_{\nu}$)}
\end{array} \right.
\end{equation} 
As shown in the right panel of Fig.~2, the resulting neutrino spectra are quite hard above PeV energies because of IR photons from the dust torus as well as internal synchrotron photons.  One of the advantages of this simple model is that the results are not sensitive to details of the blazar sequence. This is because photohadronic interactions with external radiation fields are dominant, where $f_{p\gamma}$ is not sensitive to $\Gamma$ and $\delta t'$. Also, target photon fields have narrow distributions at UV and IR bands, so that predictions for the neutrino spectral shape are reasonably robust for a given CR spectrum. 
Note that Eq.~(4) typically governs the neutrino spectral shape for HBLs, where external fields are not relevant. Even in such models, as long as the CR spectrum extends to sufficiently high energies, PeV-EeV neutrino detections are crucial to test the models.  

In general, relating the neutrino luminosity to photon luminosity is model-dependent.  For example, one can abandon the simple one-zone leptonic scenario for observed continuum spectra.  Instead, one can adopt lepto-hadronic scenarios, where gamma rays are attributed to proton synchrotron radiation or $p\gamma$-induced cascade emission, although huge CR luminosities are usually required~\cite{Sikora:2009tp,Zdziarski:2015rsa}.  In the lepto-hadronic scenarios, the relationship between the neutrino and photon luminosities is different~\cite{Petropoulou:2015upa,Padovani:2015mba}.  One of the appealing points is that the neutrino flux can be calibrated by the gamma-ray flux, and the diffuse gamma-ray intensity could be explained simultaneously with the diffuse neutrino intensity at $\sim1$~PeV~\cite{Padovani:2015mba}. 

For comparison, predictions of various blazar neutrino models are shown in the left panel of Fig.~3.  Except for an early HZ97 model, the models shown here lie in the range of the MID14 model with $\xi_{\rm cr}=3-50$.  One sees that a hard neutrino spectrum is a generic trend of the blazar neutrino models as long as a flat CR spectrum is used.  
It is possible to invoke a specific case that explains only neutrino events around PeV energies~\cite{Padovani:2015mba}, but there remains a strong tension with the absence of Glashow resonance events at 6.3~PeV.  (Note that electron anti-neutrinos come from $\pi^-$s produced via higher resonances and multi-pion production.)  
An obvious solution to reduce this tension is to introduce a spectral cutoff in the CR spectrum. This might be realized if CR acceleration is caused by stochastic acceleration rather than shock acceleration~\cite{Dermer:2014vaa}.  Explaining $\sim100$~TeV diffuse neutrinos is also possible by adopting a multi-zone emission model.  One of the physically motivated models is the spine-sheath model for the AGN jet structure~\cite{Tavecchio:2014iza,Tavecchio:2014eia}. 

The simple leptonic and lepto-hadronic scenarios of blazars have the problem that the predicted neutrino spectra are too hard to explain sub-PeV events and have tensions with IceCube upper limits above PeV energies.  Also, models are being constrained by searches for extremely high-energy neutrinos above PeV energies~\cite{Aartsen:2015zva}.
In addition, since blazars are rare objects, the absence of auto- and cross-correlation lead to strong constraints~\cite{Glusenkamp:2015jca,Wang:2015woa}, implying that their contribution to the diffuse neutrino intensity is sub-dominant especially below PeV energies. 

Note that, even if blazars are not responsible for observed diffuse neutrinos, it does not mean that they are excluded as the main sources of UHECRs.  Based on the leptonic scenario, Murase et al.~\cite{Murase:2014foa} calculated the diffuse neutrino intensity based on the hypothesis that UHECRs are produced in inner jets of radio-loud AGN (where UHECRs can be largely isotropized in bubbles, cocoons, lobes and large scale structures~\cite{Dermer:2008cy}).  The expected diffuse neutrino intensity reaches $E_\nu^2\Phi_\nu\sim{10}^{-8}~{\rm GeV}~{\rm cm}^{-2}~{\rm s}^{-1}~{\rm sr}^{-1}$ at 100~PeV energies.  Since values of $f_{p\gamma}$ for external radiation fields are more robust than those for internal radiation fields, the AGN-UHECR hypothesis can be tested in this model.  

Recently, it was claimed that a major outburst of the blazar PKS B1424-418 occurred in temporal and spatial coincidence with the 2~PeV neutrino event observed in IceCube~\cite{Kadler:2016ygj}.  The probability for a chance coincidence is $\sim5$\%, so this cannot be regarded as evidence for the blazar origin of IceCube neutrinos. Nevertheless, such the temporal and spatial coincidence can significantly reduce atmospheric backgrounds, and blazar flares are intriguing sources for high-energy neutrinos even if blazars are sub-dominant sources of the diffuse neutrino flux.  

\begin{figure}[tb]
\begin{minipage}{.49\linewidth}
\includegraphics[width=\textwidth]{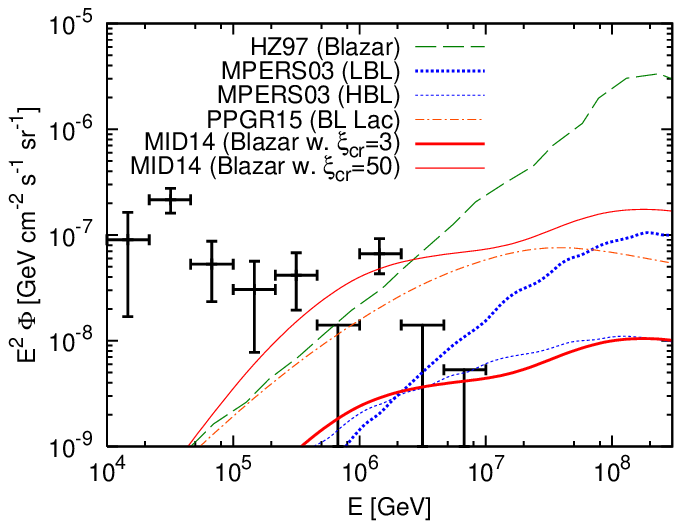}
\end{minipage}
\hfill
\begin{minipage}{.49\linewidth}
\includegraphics[width=\textwidth]{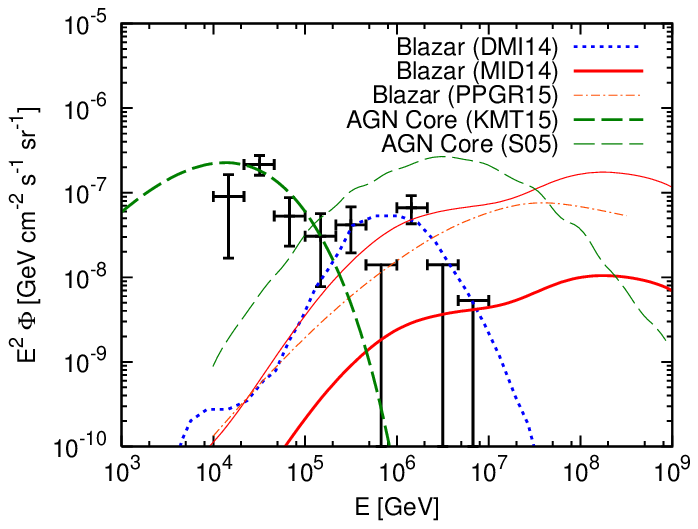}
\end{minipage} 
\caption{{\it Left panel}: All-flavor diffuse neutrino intensity calculations of various AGN jet models: (HZ97) an early blazar model by Halzen and Zas~\cite{Halzen:1997hw}, (MPERS03-LBL) a LBL model by M\"ucke et al.~\cite{Muecke:2002bi}, (MPERS03-HBL) a HBL model by M\"ucke et al.~\cite{Muecke:2002bi}, (PPGR15) a BL Lacs model by Padovani et al.~\cite{Padovani:2015mba}, (MID14 with $\xi_{\rm cr}=50$) a blazar model by Murase et al.~\cite{Murase:2014foa}, and (MID14 with $\xi_{\rm cr}=3$) a blazar model by Murase et al.~\cite{Murase:2014foa} normalized with the observed UHECR luminosity density. The diffuse neutrino intensity data from the IceCube combined likelihood analysis~\cite{Aartsen:2015knd} are also shown.    
{\it Right panel}: All-flavor diffuse neutrino intensity calculations for various AGN jet and core models: (DMI14) a FSRQ jet model by Dermer et al.~\cite{Dermer:2014vaa} normalized to the IceCube data at PeV energies assuming an average redshift $\bar{z}=2$, (MID14) a blazar jet model by Murase et al.~\cite{Murase:2014foa} with $\xi_{\rm cr}=3$ $\xi_{\rm cr}=50$ based on the leptonic scenario, (PPGR15) a BL Lacs jet model by Padovani et al.~\cite{Padovani:2015mba} based on the lepto-hadronic scenario, (KMT15) a LL AGN core model by Kimura et al.~\cite{Kimura:2014jba}, and (S05) a radio-quiet AGN core model by Stecker~\cite{Stecker:2005hn,Stecker:2013fxa}.  The diffuse neutrino intensity data from the IceCube combined likelihood analysis~\cite{Aartsen:2015knd} are also shown.
}
\end{figure}

Neutrino and gamma-ray emission from blazar jets is boosted by the relativistic beaming effect, and the corresponding high-energy emission from a single radio galaxy is expected to be weaker.  Nevertheless, the diffuse flux could be comparable because the absence of the boost for off-axis observers can be compensated by the number density~\cite{Murase:2014foa}, and nearby radio galaxies such as Centaurus A are observed at multi-wavelengths. Although the origin of TeV gamma rays is still under debate, spectral energy distributions of radio galaxies also consist of two spectral humps, suggesting an emission mechanism similar to that of blazars.  Photohadronic production of high-energy neutrinos in jets of radio galaxies has been discussed in the literature~\cite{Cuoco:2007aa,Koers:2008hv,Jacobsen:2015mga}.  However, as in the blazar case, predicted neutrino spectra are too hard to explain the IceCube data for a flat CR spectrum.    

Note that $pp$ interactions are not important in inner jets, although they could be relevant for blazar flares in some specific setups (e.g., jet-star/cloud interactions) or AGN core emission~\cite{Nellen:1992dw,Tjus:2014dna,Kimura:2014jba}.  Assuming high gas densities in the steady jets leads to serious energetics problems~\cite{Atoyan:2002gu}. Radio-loud AGN have large scale jets at kpc scales.  For such a large scale jet, observations of X-ray knots suggest column densities of $N_H\sim{10}^{20}\mbox{--}{10}^{22}~{\rm cm}^2$, implying an effective $pp$ optical depth $f_{pp}\sim{10}^{-4}~N_{H,21}$ for a jet propagating with one-third the speed of light. 
While CRs in the jet are advected, sufficiently high-energy CRs can escape from the jet and can be confined in the ambient environment~\cite{Murase:2014foa,Murase:2013rfa}.  This possibility will be discussed later.  
In addition, high-energy CRs may also be accelerated at jet-cocoon boundaries, or hot spots, cocoon shocks, and radio lobes of FR II radio galaxies.  Although they could even be relevant for UHECR production, the neutrino production there is usually inefficient~\cite{Fraija:2013goa}.

\subsection{AGN Cores}
Both radio-loud AGN and radio-quiet AGN typically show X-ray emission.  The cosmic X-ray background, which is much larger than the cosmic gamma-ray background, is known to be dominated by AGN, especially Seyfert galaxies.  Seyferts and quasars (mostly radio-quiet AGN) show so-called blue bumps at the UV band, which are naturally explained as multi-color blackbody emission from geometrically-thin, optically-thick accretion disks~\cite{Shakura:1972te}.  On the other hand, hard X-ray emission is naturally explained as Comptonized emission~\cite{hm91} by hot thermal electrons with $T\sim{10}^{9}$~K.  It is believed that hot coronae are powered by the accretion disk via e.g., magnetic reconnection.  

Although the hadronic interpretation of the observed X-ray emission is disfavored by e.g., the existence of a spectral cutoff in the spectrum, one may still consider possibilities of CR acceleration near the accretion disk.  Assuming the standard disk photon spectrum, the photomeson production efficiency for CR protons interacting with accretion-disk photons around the maximum disk temperature $kT_{\rm max}$ is estimated to be~\cite{Murase:2014foa,Dermer:2014vaa} 
\begin{equation}
f_{p\gamma}\sim600~L_{\rm AD,46.5}{(kT_{\rm max}/20~{\rm eV})}^{-1}r_{14.5}^{-1}.
\end{equation}
Thus, if CRs can be accelerated, they are efficiently depleted and high-energy neutrinos should be produced.  The typical accretion-disk temperature is $\sim10$~eV, so the neutrino spectrum resulting from $p\gamma$ interactions with the accretion-disk photons is expected to have a peak at PeV energies~\cite{Stecker:1991vm}. 
Note that $pp$ interactions are less important at high energies~\cite{Stecker:1991vm,AlvarezMuniz:2004uz} since X-ray observations indicate a column density of $N_H\sim{10}^{20}-{10}^{24}~{\rm cm}^2$.  Since CRs are depleted in the AGN core models, the neutrino flux is often normalized by X-ray and gamma-ray observations~\cite{Stecker:2005hn}, and the earlier models have overestimated neutrino production by many orders of magnitude~\cite{Achterberg:2005fs,Aartsen:2015rwa}.    

The CR acceleration mechanism is not clear in the AGN core models.  Shock dissipation may occur~\cite{brs90,AlvarezMuniz:2004uz} but CR acceleration will be inefficient when the system is radiation-dominated. When the accretion rate is high enough, protons and electrons are thermalized via Coulomb scattering within the infall time. Stochastic acceleration is unlikely in the bulk of the accretion flow, although non-thermal proton acceleration in the corona may be possible~\cite{Dermer:1995ju}.  Perhaps, electrostatic acceleration might operate, but the formation of a gap around a SMBH seems difficult except for sufficiently low-luminosity objects starved for plasma~\cite{Aleksic:2014xsg}.   Another type of AGN core model was suggested by Kimura et al.~\cite{Kimura:2014jba} for low-luminosity AGN (LL AGN).  Contrary to Seyferts and quasars, LLAGN do not have standard or slim disks, since their spectra show no blue bump~\cite{Ho:2008rf}.  Instead, their radiatively inefficient accretion flows (RIAFs)~\cite{Narayan:1994xi} are expected to be collisionless~\cite{tk85}, where particle acceleration may be possible via turbulence or magnetic reconnection~\cite{Yuan:2003dc}.  If CRs are accelerated by either stochastic acceleration or magnetic reconnection or electrostatic acceleration, high-energy neutrinos can be produced via both $pp$ and $p\gamma$ interactions~\cite{Kimura:2014jba,Khiali:2015tfa}, which may be responsible for the neutrino data in the 10-100~TeV range (see the right panel of Fig.~3).  

AGN core models may explain the IceCube data~\cite{Stecker:2013fxa,Kalashev:2015cma,Kimura:2014jba,Khiali:2015tfa}, and such hidden CR accelerators are suggested by the latest IceCube data~\cite{Murase:2015xka}.  However, all the models have large uncertainties.  Both the flux normalization and maximum energy depend on model parameters and underlying assumptions.  Although the luminosity functions, which are based on the observational data, are relatively well-known, the influence of the model uncertainties is stronger.  The number densities of radio-quiet AGN ($n_s\sim{10}^{-3}~{\rm Mpc}^{-3}$) and low-luminosity AGN ($n_s\sim{10}^{-2}~{\rm Mpc}^{-3}$) are much higher than the number densities of blazars ($n_s\sim{10}^{-7}~{\rm Mpc}^{-3}$) and radio-loud AGN ($n_s\sim{10}^{-4}~{\rm Mpc}^{-3}$).  Thus, the AGN core models are presently allowed by neutrino tomography constraints based on searches for neutrino event clustering.

\subsection{AGN in Cosmic-Ray Reservoirs}
AGN may be important as CR accelerators, even if AGN themselves are not strong neutrino or gamma-ray emitters. Radio-loud AGN are the most popular CR accelerators, and high-energy CRs and possibly UHECRs may come from jets, hot spots, cocoon shocks and lobes. Radio-quiet AGN may also have weak jets, which can also supply high-energy CRs and possibly UHECRs~\cite{Pe'er:2009rc}. In addition, disk-driven winds including ultra-fast outflows can serve as CR accelerators~\cite{Murase:2014foa,Murase:2013rfa}.  AGN are often located in galaxy clusters and groups, which have magnetic fields of $B\sim0.1-1~{\mu \rm G}$.  Low-energy CRs escaping from AGN can be confined in the large scale structure containing galaxy assemblies for $\sim1-10$~Gyr, and produce neutrinos and gamma rays~\cite{Berezinsky:1996wx}.  In this scenario, the dominant process is $pp$ interactions with intracluster or intragroup gas.  Using typical intracluster densities $\bar{n}\sim{10}^{-4}~{\rm cm}^{-3}$, with a possible enhancement factor $g\sim1-3$, we obtain~\cite{Murase:2008yt,Murase:2013rfa}
\begin{equation}
f_{pp}\simeq1.1\times{10}^{-2}~g\bar{n}_{-4}(t_{\rm int}/3~{\rm Gyr}),
\end{equation} 
where $t_{\rm int}$ is the duration of hadronic interactions.  Murase et al.~\cite{Murase:2008yt} suggested that CRs above the second knee may come from such galaxy assemblies, while CRs below the second knee are confined in the reservoirs and should produce neutrinos and gamma rays.  The case of UHECR production by AGN in clusters and groups was also studied~\cite{Kotera:2009ms}.  Interestingly, these models predicted a diffuse neutrino intensity $E_\nu^2\Phi_\nu\sim{10}^{-9}-{10}^{-8}~{\rm GeV}~{\rm cm}^{-2}~{\rm s}^{-1}~{\rm sr}^{-1}$, which may explain the high-energy neutrino data, although the contribution to the diffuse gamma-ray intensity is sub-dominant~\cite{Murase:2008yt,Murase:2009zz}.  Several authors~\cite{Murase:2008yt,Kotera:2009ms} showed a model for a central point source, where the neutrino flux is somewhat enhanced because of the higher intracluster density in the cluster/group center.  Interestingly, high-energy gamma-ray emission from the Virgo cluster center around the radio galaxy M87 can be explained by pionic gamma rays produced by interactions with the intracluster gas~\cite{Pfrommer:2013eoa}.  In reality, AGN have finite lifetimes of $\sim1-10$~Myr, and they may not be located at the center.  In the limit that CRs are injected over the whole reservoir, the neutrino spectrum is close to the injection spectrum up to the diffusion break energy, above which it becomes steeper.  Note that low mass clusters and groups, which allow us to have positive redshift evolutions, are needed for the scenario to be consistent with other gamma-ray constraints~\footnote{Although it is argued that the early work predicted much larger diffuse fluxes~\cite{Zandanel:2014pva}, actually, the early calculations for only massive clusters~\cite{Murase:2008yt,Murase:2009zz} also show $E_{\gamma}^2\Phi_\gamma\sim{10}^{-9}-{10}^{-8}~{\rm GeV}~{\rm cm}^{-2}~{\rm s}^{-1}~{\rm sr}^{-1}$, i.e., the level of diffuse gamma-ray intensities is very similar.}, including those from the gamma-ray background anisotropy and individual cluster observations~\cite{Murase:2013rfa,Zandanel:2014pva}.  Also, AGN are not the only CR accelerators in this scenario.  Not only radio-loud AGN but also radio-quiet AGN, transients in galaxies (such as supernovae and gamma-ray bursts) can contribute to the resulting neutrino and gamma-ray intensities.  

As noted above, before IceCube's discovery, Kotera et al.~\cite{Kotera:2009ms} obtained a required level of the diffuse neutrino intensity, assuming that radio-loud AGN are the UHECR accelerators. Recently, Giacinti et al.~\cite{Giacinti:2015pya} attempted to explain the gamma-ray intensity as well as the observed UHECR intensity and diffuse neutrino intensities, assuming blazars without modeling the multi-messenger emission. However, as discussed above, main blazar emission itself is unlikely to be of $pp$ origin. Non-thermal emission from radio-loud AGN including blazars is typically variable, which is most naturally attributed to inverse-Compton or perhaps $p\gamma$-induced cascade or proton synchrotron radiation.  Thus, in these scenarios, a promising possibility would be that neutrino and gamma-ray emission are mainly produced in CR reservoirs containing radio-loud AGN.  In this model, gamma rays from galaxy clusters and groups contribute to the diffuse gamma-ray background significantly~\cite{Murase:2013rfa}, and they are expected to be detected soon or have their emissivity constrained~\cite{Zandanel:2014pva,Murase:2012rd}.  

Host galaxies may also be regarded as CR reservoirs~\cite{Murase:2014foa}. However, powerful jets will leave their host galaxy, whereas weak jets or disk-driven winds from an AGN lie in the galaxy.  If CRs are accelerated by these outflows and escape from the AGN, they should interact with interstellar gas until they leave the galaxy.  Although hadronuclear production of neutrinos in radio galaxies is expected to be typically inefficient~\cite{Kimura:2014jba}, it can be important when AGN co-exist with starburst galaxies~\cite{Murase:2014foa,Murase:2013rfa}.

\begin{figure}[tb]
\begin{minipage}{.49\linewidth}
\includegraphics[width=\textwidth]{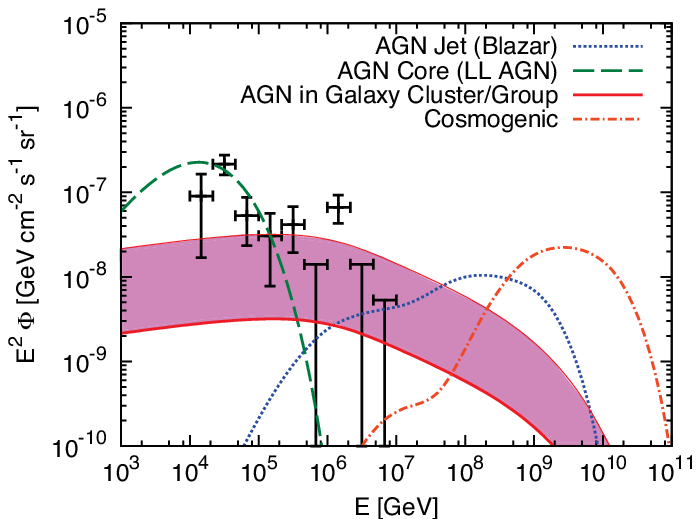}
\end{minipage}
\hfill
\begin{minipage}{.49\linewidth}
\includegraphics[width=\textwidth]{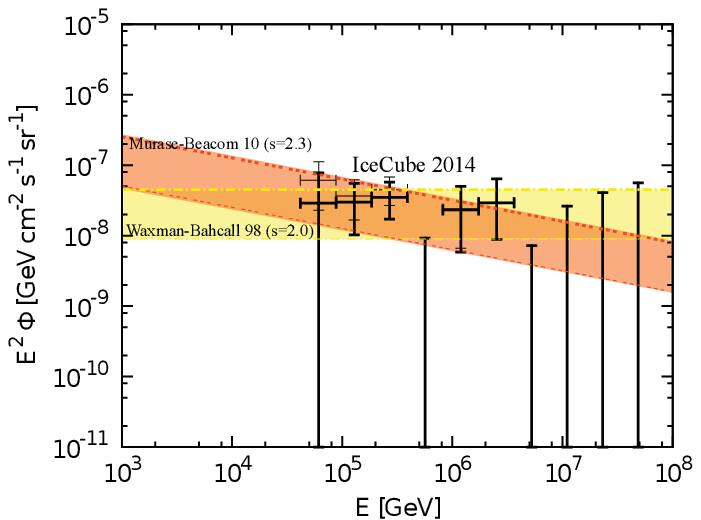}
\end{minipage} 
\caption{{\it Left panel}: All-flavor diffuse neutrino intensity calculations for various AGN-related models: (AGN Jet) a blazar model by Murase et al.~\cite{Murase:2014foa}, (AGN Core) a LLAGN core model by Kimura et al.~\cite{Kimura:2014jba}, (AGN in Galaxy Cluster/Group) a CR reservoir model by Murase et al.~\cite{Murase:2008yt} but with optimistic and moderate normalization, (Cosmogenic) a cosmogenic neutrino model by Takami et al.~\cite{Takami:2007pp} for the ankle-transition scenario.  One should keep in mind that each model has large uncertainty in its prediction.  The diffuse neutrino intensity data from the IceCube combined likelihood analysis~\cite{Aartsen:2015knd} are also shown.  
{\it Right panel}: The nucleon-survival landmark by Waxman and Bahcall~\cite{Waxman:1998yy} and nucleus-survival landmark by Murase and Beacom~\cite{Murase:2010gj}. The diffuse neutrino intensity data from the IceCube three-year high-energy starting event analysis~\cite{Aartsen:2014gkd} are also shown.
}
\end{figure}

\subsection{AGN in Intergalactic Space}
CRs escaping from AGN further interact with the cosmic microwave background (CMB) and extragalactic background light (EBL).  The most famous example is the cosmogenic neutrino production by UHECRs interacting with the CMB~\cite{Beresinsky:1969qj}.  As shown in the left panel of Fig.~4~\cite{Takami:2007pp}, cosmogenic neutrinos are expected in the EeV range, although the resulting neutrino intensity depends on the UHECR composition and redshift evolution of the sources.  The matter density in intergalactic space is so small that $pp$ interactions in cosmic voids are negligible.  Thus, for the production of PeV neutrinos, interactions with the EBL in the UV range are relevant, and the photomeson production efficiency is estimated to be
\begin{equation}
f_{p\gamma}\approx \hat{n}_{\rm EBL}\sigma_{p\gamma}^{\rm eff}d\simeq1.9\times{10}^{-4}~\hat{n}_{\rm EBL,-4}d_{28.5},
\end{equation}
where $\hat{n}_{\rm EBL}\sim{10}^{-4}~{\rm cm}^{-3}$ is the number of EBL photons and $d$ is the particle travel distance. Since the efficiency is tiny, for intergalactic neutrino production to be relevant, optimistic EBL models and large CR luminosity densities are required~\cite{Kalashev:2013vba}.  As in the blazar case, one has to optimize the CR maximum energy not to overproduce neutrino events above a few PeV energies, and the intergalactic origin of sub-PeV neutrinos is unlikely~\cite{Roulet:2012rv,Laha:2013eev}.

\section{Discussion and Summary}
Radio-loud AGN have powerful jets, which are promising CR accelerators. The spectral energy distributions and luminosity function of blazars have been measured reasonably well.  For power-law CR injections, most blazar neutrino models predicted hard neutrino spectra and the peak energy is expected in the 10-100~PeV range.  The absence of neutrino events at multi-PeV energies and the lack of auto and cross correlations imply that the simple jet models of radio-loud AGN are already disfavored as the main origin of the observed diffuse neutrinos.  More complicated scenarios may be necessary.  Possibly, only neutrino events around $\sim1$~PeV could be explained by blazars~\cite{Dermer:2014vaa,Padovani:2015mba}.  Or one can invoke multi-zone emission models and/or introduce non-power-law CR spectra such as a log-parabolic function motivated by stochastic acceleration~\cite{Dermer:2014vaa,Tavecchio:2014iza}.  

However, blazars do not have to be dominant sources of the observed neutrinos in IceCube. They may produce very-high-energy neutrinos without explaining the sub-PeV neutrinos because of their hard neutrino spectra. Indeed, some models~\cite{Murase:2014foa} predict that, in addition to whatever produces the IceCube neutrinos, there might also be a low level of very-high-energy neutrinos from blazars that become prominent above a few PeV energies.
In my personal view, searches for $10-100$~PeV (or higher-energy) neutrinos with IceCube, KM3Net, IceCube-Gen2~\cite{Aartsen:2014njl}, ARA~\cite{Allison:2011wk}, ARIANNA~\cite{Barwick:2006tg} and GRAND~\cite{Martineau-Huynh:2015hae}, seem more interesting.  Improving sensitivities in this very-high-energy energy range will allow us to constrain a significant part of the parameter space of various blazar neutrino models.  In particular, their connection to UHECRs can be critically examined.  Apparently, the observed diffuse neutrino intensity is compatible with the Waxman-Bahcall bound for a spectral index $s=2.0$~\cite{Waxman:1998yy}.  However, in the blazar neutrino models, a flat CR spectrum leads to a hard neutrino spectrum since $f_{p\gamma}$ increases with energy, so the simultaneous explanation of the observed UHECR and neutrino intensities is difficult~\cite{Murase:2014foa,Kistler:2013my,Yoshida:2014uka}.  Nevertheless, steeper CR spectra might help.  Indeed, the observed diffuse neutrino intensity is also compatible with the nucleus-survival bound for a spectral index $s=2.3$~\cite{Murase:2010gj} (see the right panel of Fig.~4). Note that, for blazars to produce UHECRs, the composition is expected to be heavier at ultrahigh energies. Inner jets of FSRQs and FR II galaxies could accelerate protons to $\sim{10}^{20}$~eV.  However, for such luminous blazars, the IR dust-torus component can deplete UHECR protons and neutrons~\cite{Murase:2014foa}.  Since the photodisintegration cross section is higher, this is even the case for heavy nuclei.  On the other hand, inner jets of BL Lacs and FR I galaxies are not powerful enough to accelerate protons to $\sim{10}^{20}$~eV~\cite{Murase:2011cy}, especially in the leptonic model for gamma rays.  As a result, the AGN-UHECR hypothesis in the simple model predicts that $\sim10-100$~PeV neutrinos large come from FSRQs while UHECRs come from BL Lacs and FR I galaxies.     

Not only radio-loud AGN but also radio-quiet AGN can be the sources of diffuse neutrinos. For Seyferts and quasars, if CRs are accelerated in the vicinity of a SMBH, efficient $p\gamma$ interactions with UV and X-ray photons from the standard accretion disk and corona are expected.  Alternatively, LL AGN associated with RIAFs have been considered as potential non-thermal particle emitters, and high-energy CRs may be accelerated by turbulence or magnetic reconnection.  Although it is possible for the models to fit the IceCube data, model uncertainties are quite large at present and further theoretical and observational studies may be needed. 

In contrast, CR reservoir scenarios have a strong predictive power.  In this work, we discussed CR reservoir scenarios involving AGN, which are different from the starburst model~\cite{Murase:2013rfa}. Both radio-loud and radio-quiet AGN embedded in such reservoirs may contribute to the observed diffuse neutrino intensity, and a spectral break due to CR diffusion was expected before IceCube's discovery~\cite{Murase:2008yt,Kotera:2009ms}.  
Although a part of the parameter space has been constrained by multi-messenger data, it is appealing that the connection between observed CRs above the second knee (that may include UHECRs) and PeV neutrinos is expected in this model.  The contribution to the diffuse gamma-ray background would be sub-dominant~\cite{Murase:2013rfa}.  
 
AGN are widely considered as promising CR accelerators and neutrino sources.  However, many problems related to their non-thermal activities remain unresolved.  High-energy neutrinos have provided us with a new probe of the physics of AGN, and detailed comparisons to various theoretical models have been made possible.  I hope that further multi-messenger studies will help us solve some of the mysteries, especially the long-standing question whether AGN are the sources of UHECRs.  

K.M. acknowledges Chuck Dermer, Tom Gaisser, Gwenael Giacinti, Shigeo Kimura, Peter M\'esz\'aros, Foteini Oikonomou, Maria Petropoulou, Shigeru Yoshida, and Fabio Zandanel for discussion and comments. 

  
\end{document}